\documentclass[
twocolumn,
prl,showpacs,floats,preprintnumbers,amsmath,amssymb]{revtex4}
\usepackage{epsfig} 
\usepackage{amsmath}

\renewcommand{\[}{\left [}
\renewcommand{\]}{\right ]}
\def\fslash#1{#1 \!\!\! \slash}
\def\beq{\begin{equation}}
\def\eeq{\end{equation}}

\def\varp{\varepsilon}
\def\bea{\arraycolsep .1em \begin{eqnarray}}
\def\eea{\end{eqnarray}}

\def\vp{{\bf p}}

\def\vk{{\bf k}}

\def\Tr{{\rm Tr}}

\let\si=\sigma

\let\om=\omega   
    
\let\no=\nonumber

\def\eq#1{Eq.(\ref{#1})}
\def\refr#1{\cite{#1}}

\def\s0#1#2{\mbox{\small{$ \frac{#1}{#2} $}}}
\def\0#1#2{\frac{#1}{#2}}

\def\anp#1#2#3{Adv.\ Nucl.\ Phys. \ {\bf #1}, #2 (#3)}
\def\plb#1#2#3{Phys. Lett. {\bf B #1}, #2 (#3)}
\def\npa#1#2#3{Nucl. Phys. {\bf A #1}, #2 (#3)}

\def\prc#1#2#3{Phys. Rev.  {\bf C #1}, #2 (#3)}

\def\prv#1#2#3{Phys. Rev. {\bf #1}, #2 (#3)}

\def\ann#1#2#3{Ann. Phys. {\bf #1}, #2 (#3)}
\def\anp#1#2#3{Adv. Nucl. Phys. {\bf #1}, #2 (#3)}

\def\jhep#1#2#3{J.\ High Energy Phys.\ {\bf #1}, #2 (#3)}
\def\zpa#1#2#3{Z.\ Phys.\ {\bf A #1}, #2 (#3)}
 
\def\ibid#1#2#3{{\it ibid.}, {\bf #1}, #2 (#3)}
\begin{document}

\title{
$^1 S_0$ pairing correlation in symmetric nuclear matter 
with Debye screening effects}
\author{Ji-sheng Chen\thanks{chenjs@iopp.ccnu.edu.cn}$^{a}$\footnote{Email
address: Chenjs@iopp.ccnu.edu.cn},~~~~~~Peng-fei
Zhuang$^{b}$,
and Jia-rong Li$^{a}$
}
\affiliation{$^a$
Physics Department, Hua-Zhong Normal University, Wuhan 430079,
People's Republic of China\\
$^b$Physics Department, Tsinghua University, Beijing 100084, People's
Republic of China}
\begin{abstract}
The $^1 S_0$ pairing of symmetric nuclear matter is discussed in the frame
	work of relativistic nuclear theory with Dyson-Schwinger equations(DSEs).
The in-medium nucleon and meson propagators are treated in a more
	self-consistent way through meson polarizations. 
The screening effects on mesons due to in-medium nucleon excitation 
	are found to reduce the $^1S_0$ pairing gap and shift remarkably the gap peak  to 
	low density region.
\end{abstract}
\maketitle

Compared with nonrelativistic frame work, the relativistic nuclear theory
	can successfully describe the saturation at normal nuclear density. 
The basic meson exchange is normally considered as the nuclear saturation mechanism. 
The original $\sigma-\om $ theory of quantum hadrodynamics model (QHD) 
	developed by Walecka et al. and its various extensions have been widely used
	to discuss the properties of
	finite nuclei and nuclear matter\refr{walecka1974,serot1986,boguta1977,zm1990}.

Superfluidity of strongly interacting Fermi system is very important 
	for understanding the properties of
	finite nuclei, such as the dramatic reduction of the moments of inertia in
	rotating nuclei or the energy gap in the spectra of many even-even nuclei\refr{ring1980,bohr1958}.
The existence of superfluidity may also affect the dynamical and thermal evolution
	mechanism of neutron stars because it is closely related to the emission of 
	neutrino and cooling of neutron-rich matter.
It is also argued that the superfluidity of nuclear matter can lead to
	the glitches of astronomy phenomena and attracts much attention in
contemporary physics\refr{glendenning2000}.  

Although there are many works in the literature 
	addressing the superfluidity of nuclear matter, 
	the main results are obtained from the
	non-relativistic nuclear theory and no definite conclusion 
	can be made yet.
We noted that since K. Kucharek and P. Ring\refr{ring1991} first derived the relativistic Hartree-Fock-Bogoliubov
	equation by using Green function method and the Gor'kov factorization
	analogously to nonrelativistic BCS theory, it was found that the
	superfluidity gap value is about three times larger than the ``standard" value obtained
	with the nonrelativistic Gogny force\refr{gogny1980}. 
To improve the description of
	superfluidity with relativistic nuclear theory, 
	various	approaches have been investigated, 
	such as using external potential as input or various cut-offs of the integration 
	momentum\refr{matsuzaki1998,matsuzaki1999,serra2001}.
To our knowledge, there is even no definite result about the superfluidity gap(the
	gap value and peak position) for symmetric nuclear
	matter within nonrelativistic or relativistic nuclear theory up to now.
However, it is widely accepted that the $ ^1S_0$ gap values at normal nuclear
	density should be very small.

Essentially one can not expect that the softness of equation of state(EOS) 
	describing the bulk property of nuclear matter is directly related with
	the superfluidity property of nuclear matter.
For example, although the nonlinear $\sigma$-$\omega$
	model with the possible embarrassing negative
	coupling constants $b$ and $c$ (which in principle lead to instability of nuclear system
	at high density scenario) in $\sigma $ self-interaction terms $b\sigma ^3+c\sigma ^4$ 
	can give a very soft EOS with mean field theory(MFT)\refr{glendenning2000}, 
	the gap behavior is similar to that in the original version 
	by using frozen meson propagators\refr{ring1991,serra2001}.

\begin{figure}[ht]
	\centering
	\psfig{file=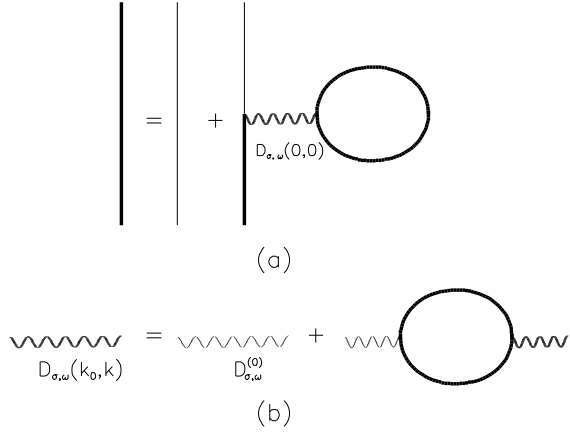,width=7.5cm,angle=-0}
	~\\[.2cm]
	\caption{
	\small Diagrammatic representations for the propagators of in-medium nucleons(a) and
	mesons(b).
	}\label{fig1}
\end{figure}

Theoretically, since it is difficult to make low energy calculations directly with 
	quantum chromodynamics(QCD), one has to work with
	effective theories.
As an effective theory, QHD-I model (and its extensions) has been widely
	used to discuss the effective meson masses under extreme environment in the
	past\refr{chen3,shiomi1994,chen2002,chen2003}.
In principle, when one discusses the in-medium properties
	of nucleons and mesons, one has to take into account
	the back-interactions of nucleons with in-medium mesons.
Therefore the {\em resummed} nucleon and meson propagators would form a closed set of
	coupled equations and should be solved simultaneously.
With this self-consistent way, a softer EOS with an
	acceptable compression modulus ${\cal K}$ in dealing with realistic nuclear
	matter can be obtained\refr{Bhattacharyya1999,chen3}.
In the spirit of mean field theory, the exchanged mesons in determining nucleon propagator
are not free but medium dependent.
Their masses should be determined together with the nucleon mass through
	 Dyson-Schwinger equations self-consistently, as indicated by Fig.\ref{fig1}.

\begin{figure}[ht]
	\centering
	\psfig{file=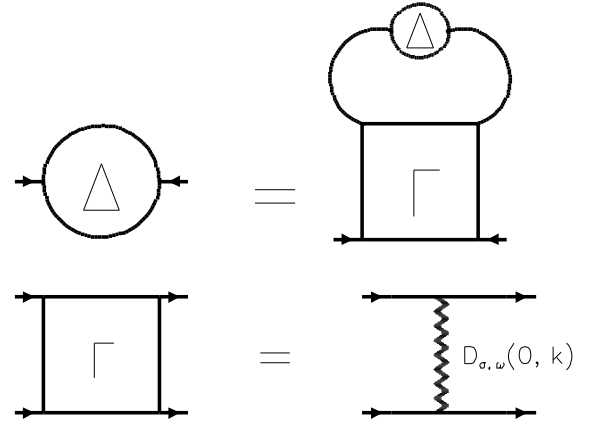,width=7.5cm,angle=-0}
	\caption{
		\small Diagrammatic representations for the gap equation and the interaction
		kernel in instantaneous
		approximation with {\it screened} meson propagator. 
	}\label{fig2}    
\end{figure}

It would be very interesting to analyze the in-medium effect of mesons on the
	superfluidity of nuclear matter in the frame work of relativistic nuclear theory. 
In superfluidity state and with QHD-like Lagrangian,  
   Dyson-Schwinger equations for the nucleon and meson propagators  as indicated in Fig.\ref{fig1}
     and the energy gap equation 
           as indicated in Fig.\ref{fig2} form a new closed set of coupled equations.   
The in-medium meson propagators $D$ instead of the normally used bare ones $D^{(0)}$ 
     will affect the kernel in the BCS gap equation.
We will see below that the gap behavior with in-medium meson propagators is 
	quite different from that with bare ones.
We found that the polarization effects leading to screening have been
	widely discussed in the nonrelativistic frame work of nuclear
	theory\refr{bahu1973,clark1976,bozek1999,baldo2000,schulze2001,lombardo2001}, 
	this problem has not yet been addressed
	before in terms of relativistic field theory. 
In this letter, we want to discuss the effects of in-medium effective potential
	for nucleon-nucleon interaction on the superfluidity due to
screening by using the original
	renormalizable formalism of $\si-\om $ model.

Let us start with the four-dimensional gap equation by using the standard
Nambu-Gor'kov formalism in the ladder approximation of the meson exchanges as indicated
	by Fig.\ref{fig2}\refr{schulze2001,gorkov1963,schrieffer1964},
\bea
	\Delta ^* (K) =i \int \0{d^4P}{(2\pi)^4}<P|\Gamma |K> F^+ (P)
\eea
where $K$ is the four momentum $K=(k_0,\vk)$, $<P|\Gamma |K>$ is the interaction
kernel and  
\bea
	F^+(K)=\0{-\Delta ^* (K)}{\[k_0-\varp (K) +i \eta \]\[k_0-\varp (-K) -i \eta
	\]-|\Delta (K)|^2}\no
\eea 
	is the Nambu-Gor'kov anomalous propagator with 
	$\varp (K)=E_k-E_{k_f}$ being the quasi-particle energy above Fermi-surface. 
For $^1S_0$ pairing, the gap equation can be reduced to\refr{ring1991,matsuzaki1998}
\bea\label{gap_3}
	\Delta (p)=-\0{1}{8\pi^2 }\int _0^\infty {\bar v}_{pp}(p,k)\0{\Delta (k)
	}{\sqrt{(E_k-E_{k_f})^2+\Delta ^2 (k)}} k^2 dk,\no\\
\eea 
	with $E_k=E_k^*+\lambda $ and $E_k^*=\sqrt{M_N^{*2}+\vk ^2}$. 
The quantity $\lambda$ related with the baryon current
	is obtained from the tadpole self-energy of 
	nucleon propagator with in-medium vector meson in Fig.\ref{fig1}
\bea\label{five}
	\lambda =\0{g_\om ^2}{\bar{m}_\om ^2}\0{\gamma}{2\pi^2}\int _0^\infty
	v_k^2 k^2 dk,
\eea
	where $\gamma=4$ is the spin-isospin degeneracy factor for symmetric nuclear
	matter, and $v_k^2 $ is the BCS occupation number
\bea
	v_k^2=\012(1-\0{E_k-E_{k_f}}{\sqrt{(E_k-E_{k_f})^2+\Delta ^2(k)}}).\eea

The interaction kernel $<P|\Gamma|K>$
	in our treatment is approximated by
	the in-medium meson propagators instead of the bare ones. 
We use the static(instantaneous) approximation by neglecting the retarding effects\refr{ring1991}. 
Since the meson propagators
	 $D_{\sigma, \om} (0, \vk )$ with the vanishing temporal component of
	four-momentum are now medium dependent, 
	the Debye screening effects will play an important
	role in the in-medium particle-particle interaction potential
\bea
	&&\bar{v }(\vp,\vk)=<\vp s', {\widetilde{\vp s'}}|V|\vk s, {\widetilde{\vk
	s}}>-<\vp s', {\widetilde{\vp s'}}|V|{\widetilde{\vk    
	s}, \vk s}>\no\\
	&&~~~=\mp \0{M_N^{*^2}}{2 E^*(k) E^*(p)}\0{\Tr [\Lambda _+ (\vk ) \Gamma \Lambda
	_+ (\vp )\gamma ^0
	{\cal T}^+ \Gamma  ^+{\cal T}\gamma^0]}{(\vk -\vp )^2 +m_D ^{*^2}},
\eea
where $\Lambda _+ (\vk )=\0{\fslash{k}+M_N^*}{2M_N^*}$ is the projection operator of the positive
energy solution and ${\cal T}=i \gamma^1 \gamma ^3$ is the time reversal
operator.  
The tilde typescript ``$\sim $ " means the Gor'kov time reversal state
	and $\Gamma $ is the corresponding interaction vertex of $\sigma $/$\omega $
	with nucleons.
The assymetrized matrix elements $\bar v_{pp} (p,k)$ in
	the gap equation \eq{gap_3} for $^1S_0$ pairing is obtained through the integration of
	$\bar{v}(\vp ,\vk ) $ over the angle
	$\theta $ between the three-momentums $\vk $ and $\vp$
\bea
	\bar v_{pp} (p,k)=\int \bar{v}(\vp ,\vk ) d \cos\theta .
\eea
The effective nucleon mass $M_N^*$ is determined by the 
	relevant mass gap equation through tadpole self-energy of nucleon propagator
with scalar meson
\bea\label{four}
	M_N^* =M_N-\0{g_\sigma ^2 }{\bar m_\sigma ^2}
	\0{\gamma}{2 \pi^2}
	\int _0^\infty
	\0{M_N^*}{E^*_p}v_p^2 p^2 dp+\Delta M^*_{N,vac}, 
\eea 
with $\Delta M^*_{N,vac}$ being the vacuum fluctuation
contribution
\bea
	\Delta M^*_{N,vac}&&=\0{g_\sigma^2}{{\bar m}_\sigma^2}\01{\pi ^2} \[M_N^{*3}\ln
	(\0{M_N^*}{M_N})-M_N^2 (M_N^*-M_N)~~~~
	\right.\no\\&&\left.    
	-\052 M_N (M_N^*-M_N)^2-\0{11}6
	(M_N^*-M_N)^3\].
\eea

The polarization tensors $\Pi _{\sigma ,\omega}  (k_0,\vk)$
	determining the in-medium $\sigma $ and $\omega$ propagators are
	calculated by using corresponding Dyson-Schwinger equations as	shown in
Fig.\ref{fig1}. 
For brevity, here we list only the sigma meson self-energy explicitly,
\bea
\label{si}  
	&&\Pi_ \sigma(k)=\frac{3 g_{\sigma}^2}{2 \pi^2} \left [3M_N^{*^2} -4 M^*_N
	M_N+M_N^2 
	\nonumber\right. \\&&\left.
	-(M_N^{*^2} - M_N^2) \int_{0}^{1}
	\ln\frac{M_N^{*^2}
	- x(1 - x)k^2}{M_N^2} dx \nonumber\right. \\
	  &&\left.- \int_{0}^{1} (M_N^2 - x(1 - x)k^2) \ln\frac{M_N^{*^2} - x(1 -x)k^2}
	{M_N^2 - x(1 - x)k^2} dx\right ]
	\no\\&& ~~~~~~
	+\0{g_\sigma ^2}{\pi ^2}\int _0^\infty  \0{v_p^2 p^2 dp }{E_p^*}\[2+\0{k^2-4 {M^*_N}^2  }{4 p |\vk|}(a+b
	)\],
\eea
with
\bea
	a=&&\ln\0{k^2-2 p |\vk|-2 k_0 E^*_p }{k^2+2 p |\vk|-2 k_0 E^*_p},~~~
	b=a(E^*_p\rightarrow -E^*_p).\no
\eea
  
The effective masses ${\bar m_\sigma }$ and  ${\bar m_\omega }$ in \eq{five} and \eq{four}  
	are determined by the corresponding polarization tensors with vanishing four-momentum transfer,
	and the Debye screening masses $m_\si ^*$ and $m_\om ^*$ in
	assymetrized matrix elements $\bar{v}_{pp} (p,k)$  
	are determined by the pole positions of 
	corresponding spacelike propagators $D_{\sigma,~\om} (0,\vk)$ due to taking the
	static approximation\refr{chen2003}. 
For example, the transverse mode screening mass $m^*_\om $ is determined self-consistently by 
\bea
	m^{*2}_\om=m^{2}_\om +\Pi ^T_\om (0, i m_\om^*),
\eea
where $\Pi ^T_\om $ is the transverse part of polarization tensor $\Pi _\om^{\mu\nu}(k_0,\vk) $.
In principle, the longitudinal mode screening mass is different from the transverse mode
	one for in-medium vector meson due to the broken Lorentz invariance.
However, neglecting this little
	difference doesn't affect the qualitative result in realistic numerical calculation.

Considering the in-medium meson effects on the property of nuclear matter,
	one should refix the parameters in the model. 
Noting that the effect of superfluidity
	on the bulk property is negligible, we fix the parameters by normal nuclear matter with saturation
	condition of binding energy $e_n=-15.75$ MeV at the normal nuclear density with $k_f^0=1.42 fm^{-1}$. 
The relevant parameters are listed in Table.\ref{tab}. 

\begin{figure}[ht]
	\centering
	\psfig{file=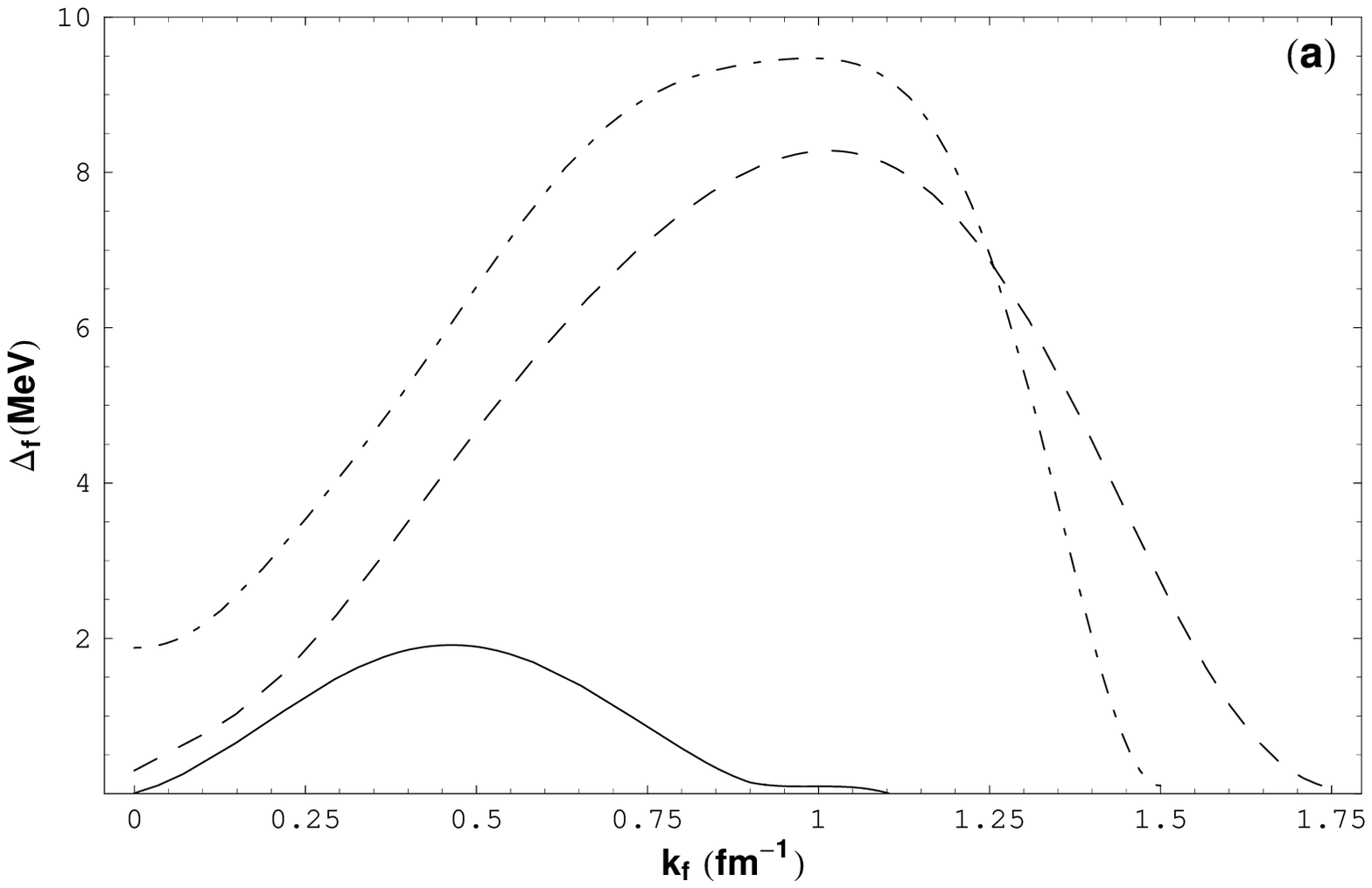,width=7.5 cm,angle=-0} ~\\[.2cm]
	\psfig{file=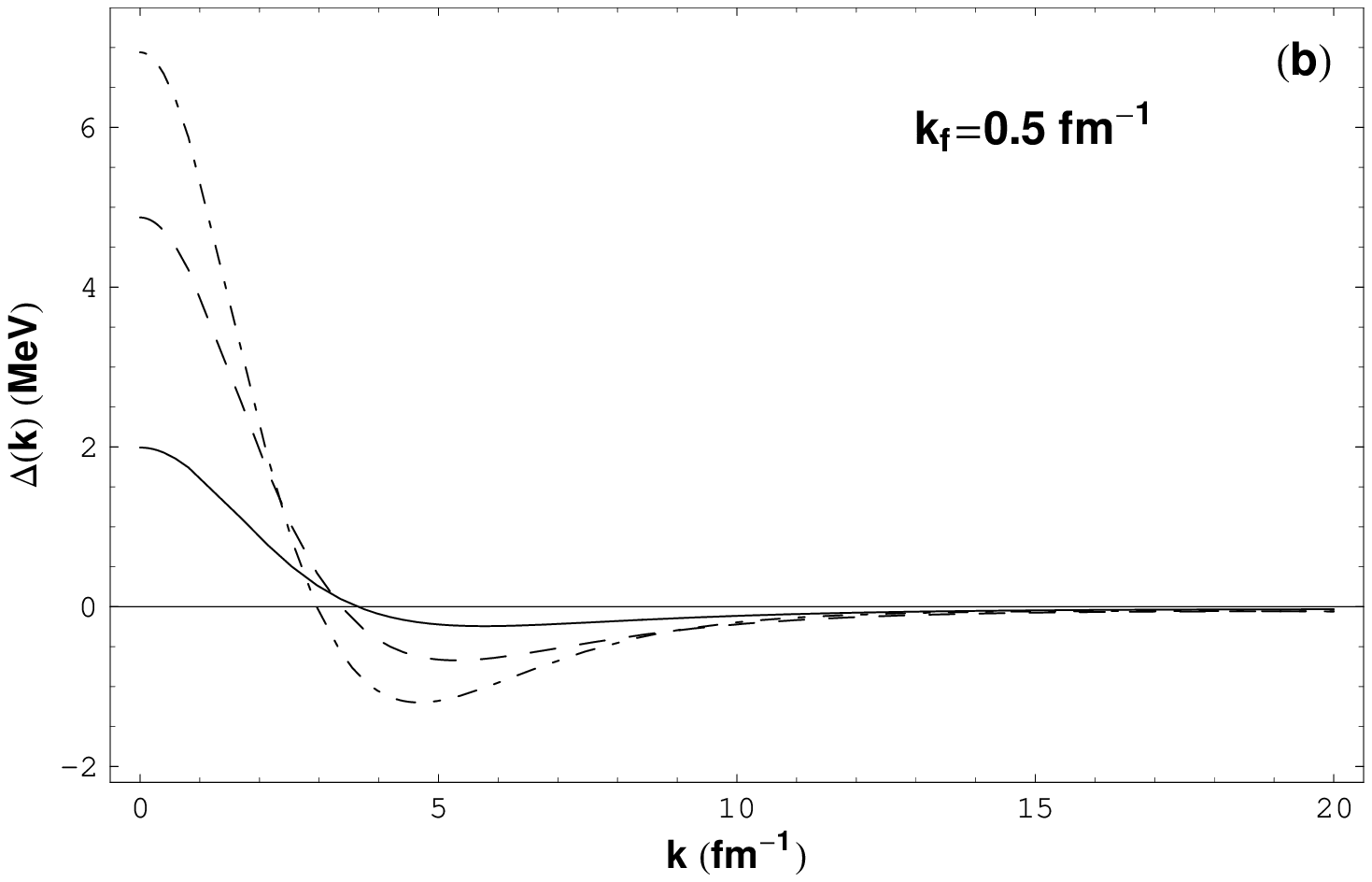,width=7.5 cm,angle=-0}
	\caption{
	 (a) The pairing gap $\Delta _f$ at the Fermi surface as a function of density
	characterized by the Fermi momentum $k_f$. 
	 (b)
	The gap function $\Delta (k)$ as a function of momentum k for
	fixed Fermi momentum $k_f=0.5 fm^{-1}$. 
	The dot-dashed lines correspond to
	the result obtained by MFT and dashed lines correspond to
	RHA\refr{matsuzaki1998}, while the solid lines correspond to our
	self-consistent resummation approach. }
\label{fig3}    
\end{figure}

The remaining task will be the numerical solution of the coupled equations indicated by Figs.\ref{fig1} and
	\ref{fig2}. 
It should be noted that the relativistic kinematic factors guarantee the
	convergence of the gap equations such as \eq{gap_3}  for the relativistic nuclear theory and
	lead to a definite result for the gap. 
In principle, the momentum integration upper bound in relevant equations such as
	in \eq{gap_3}  is infinity. 
However, a concrete upper bound must be used to give a numerical result by solving the gap integral equation.
Strictly speaking, the gap value should not be sensitive to the adopted
	momentum upper bound and 
	the sensitivity of momentum cut-off on the gap has been analyzed in such as in Refs.\refr{ring1991,matsuzaki1998},
	which can be also reflected by the gap function indicated by Fig. \ref{fig3}(b).  
A concrete and large enough momentum upper bound $\Lambda _p=20 fm^{-1}$ has been used in this work
for the description of screening effects.
To focus on the characteristic due to polarization effects,
	the $\sigma-\omega$ mixing effects have been neglected,
	which will not affect the result qualitatively although it deserves
	further study.

The numerical results of the superfluidity gap equations are shown in
Fig.\ref{fig3}.
In the upper panel (a), we indicate the gap curves $\Delta (k_f)$ versus Fermi momentum $k_f$, 
and the gap functions $\Delta(k)$ at given Fermi surface momentum $k_f =0.5 fm^{-1}$  
are shown in Fig. \ref{fig3}(b).

\begin{figure}
	\centering
	\psfig{file=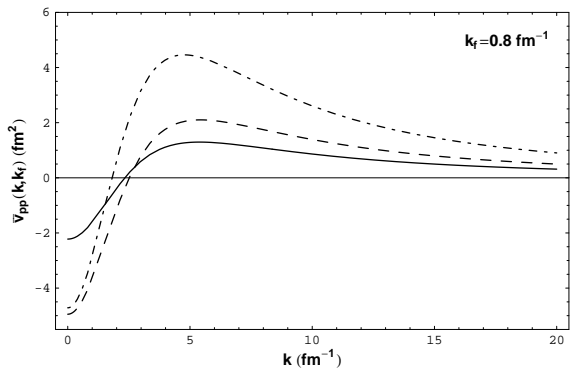,width=7.5 cm,angle=-0}
	\caption{ \small 
	Assymetrized matrix elements $\bar{v}_{pp} (k,k_f)$ in momentum space  at the Fermi
	momentum $k_f=0.8 fm^{-1}$. The line-styles are similar to those in
Fig.\ref{fig3}.
	}\label{fig4}
\end{figure}

Compared with the previous superfluidity results of relativistic nuclear theory in the literature, 
	the gap value we obtained 
	is very small and the peak position is shifted to the 
	low density region remarkably.
As mentioned in the introduction, this 
	interesting result is not due to the softener EOS but much
	attributed to the screened effective
	particle-particle interaction potential as indicated by Fig.\ref{fig4}.
The key point is that the $\sigma $ and $\omega $ propagators in the gap equations are not
	bare but in-medium ones determined self-consistently by Dyson-Schwinger equations. 
The effective nucleon-nucleon interaction potential
	with Debye screening of in-medium mesons
	leads to the change in the interaction force range.
Different from the scenario of bare meson propagators used in the gap equation, 
	the particle-particle potential is more sensitive to
	density, which can be understood from the corresponding attractive and repulsive
	force range changes for different densities characterized by Fermi momentum $k_f$(not
	displayed obviously in Fig. \ref{fig4}). 

It is clearly demonstrated in Fig.\ref{fig3} that our self-consistent approach
	reduces the difference
	between the nonrelativistic and relativistic theories about the maximum gap
	value and the peak position of $^1S_0 $ pairing correlation.
The significant improvement by the self-consistent resummation approach for
	the particle-particle interaction leading to pairing is reflected on two
	aspects: One is at the saturation density with $k_f^0=1.42 fm^{-1}$, the
	other is at $k_f=0$. 
The improvement at $k_f=0$ is crucial by noting that the MFT and RHA approaches with frozen meson propagators give unrealistic
	{\em non-zero} gap values $\sim
	1.94$ MeV/$0.36$ MeV\refr{matsuzaki1998}.

Summarizing, with a set of more self-consistent equations for the 
	resummed in-medium nucleon and meson propagators and the superfluidity gap
	by Dyson-Schwinger Green function approach, 
	we have studied the $^1S_0$ pairing correlation in symmetric nuclear matter and compared our results with those 
	obtained by MFT and RHA approaches.
The Debye screening effects of in-medium meson propagators
can reduce significantly  the superfluidity gap value, while the gap peak position is
shifted remarkably to low density region. 

{\bf Acknowledgments: }{One of the authors(Ji-sheng Chen) acknowledges the beneficial communications and
valuable discussions with M.~Serra and M.~Matsuzaki through Prof. P. Ring. This work was supported by NSFC
under Nos. 10135030, 10175026, 19925519, 90303007 and
the China Postdoc Research Fund.}
\begin{table}
\caption{
The parameters determined at normal nuclear density with $k_f^0=1.42 fm^{-1}$ in MFT, RHA 
	and our self-consistent approach(labeled as SA)
 	with $M_N=939$ MeV,
	$m_\om =783$ MeV and
	$m_\si =550$ MeV.  
We show also the compression modulus $\cal K$ (in MeV), 
 the {\em medium dependent} coupling constants
 (determining EOS) $C_s^2=g_\si^2$$\0{M_N^2}{\bar m_\si^2}$ and $C_v^2=g_\om^2$
 $\0{M_N^2}{\bar m_\om^2}$, 
 the maximum of gap value $ \Delta^m_f$ (MeV), 
the peak position $k^m_f$ ($fm^{-1}$) and the ``gap value" $\Delta (0)$ (MeV) at $k_f=0$.
}
\begin{ruledtabular}
   \begin{tabular}{cccccccccc}
     & $g_\sigma^2$ & $g_\om^2$  & $C_s ^2$&$C_v^2$& ${\cal K}$&$\0{M_N^*}{M_N}$&$\Delta^m_f$&$k^m_f$&$\Delta(0)$ \\
     \colrule
	MFT &91.64&136.2 &267.11&195.87&545.43&0.556&9.4&1.0&1.94\\
     \colrule
	RHA &62.89&79.78  &183.31&114.73&468.24&0.718&8.3&1.0&0.36  \\
     \colrule
	SA &48.90 &53.40&123.17&66.078 &338.00&0.794&1.9&0.5&0 \\
   \end{tabular}
\end{ruledtabular}
 \label{tab}
\end{table}

\end{document}